# Magnetically tunable topological singularity of Moiré bound states in the continuum

Shuo Wang[1†], Sheng-Yi Wang[1,2†], Jie Yang[3], Wenkui Zhao[1], Haifeng Kang[2], Qiu Wang[1], Song Han[4,5], Bo-Wen Jia[1*]

[1]*School of Information Engineering and the Hubei Key Laboratory of Broadband Wireless Communication and Sensor Networks, Wuhan University of Technology, Wuhan 430070, China*

[2]*Key Laboratory of Artificial Micro- and Nano-structures of Ministry of Education, and School of Physical and Technology, Wuhan University, Wuhan 430072, China*

[3]*School of Physics, Huazhong University of Science and Technology, Luoyu Road 1037, Wuhan, 430074, China*

[4]*Innovative Institute of Electromagnetic Information and Electronic Integration, College of Information Science &Electronic Engineering, Zhejiang University, Hangzhou, 310027, China*

[5]*State Key Laboratory of Extreme Photonics and Instrumentation, ZJU-Hangzhou Global Scientific and Technological Innovation Center, Zhejiang University, Hangzhou 310200, China*

† These authors contributed equally to this work.

* Email: jiabowen@whut.edu.cn

## Abstract:

This work investigates the dynamic manipulation of band structure and the topological polarization singularities of magneto-optical Moiré bound states in the continuum (BICs) formed by superimposing two conventional magneto-optical BIC gratings with slightly mismatched periods. By controlling the mutual displacement of two conventional BIC gratings to break the local geometric symmetry of Moiré superlattice, the topological charge can be split into two half integer topological charges with opposite chirality while the Q factor maintain much higher value than conventional BICs. By applying an external magnetic field oriented in various directions to further break the time-reversal symmetry, we demonstrate dynamic control over the Moiré BICs' topological singularities. Notably, intrinsic optical chirality emerges when both mirror and time-reversal symmetries are broken simultaneously. The resulting C points can be tuned by adjusting the magnetic field strength and the geometric parameters of the Moiré pattern. Our findings reveal that the topological singularities of Moiré BIC are flexibly engineerable by external fields and that maximal optical chirality occurs under concurrent symmetry breaking. This work offers a new approach to realizing tunable topological radiation and chiral photonic responses in Moiré photonic crystals, with potential applications in enhancing chiral-optical effects and improving the tunable performance of Moiré optoelectronic devices.

## I. INTRODUCTION

Topological singularities [1-3] play a significant role in photonic systems, particularly in micro- and nano-photonics, where they are intrinsically associated with the fundamental properties of BIC [4-6]. These singularities manifest as polarization vortices in momentum ($k$) space, each characterized by topological charges [7-8]. These integer-valued charges originate from specific interactions between light and materials [9-11]. The ability to dynamically manipulate these topological singularities is crucial for controlling optical modes and polarization states, enabling advanced photonic applications, such as ultrasensitive biosensors [12-13] and low-threshold lasers [14-15]. Dynamic modulation of topological singularities, particularly shifts and splits of topological charges in $k$ space, is essential for adjusting the optical properties of BIC [16-18]. For example, changing the geometric structure of photonic crystals (PhCs) allows for the repositioning of topological singularities in $k$ space, consequently tuning the associated topological charges and optical behaviors [19]. By varying the lattice constant of a PhC slab, two off-Γ BIC bearing identical topological charge can converge at the Γ point [20]. Similarly, altering the slab thickness can induce multiple BIC to merge into a single higher-order singularity at the Γ point [21].

Furthermore, external fields have emerged as powerful tools to dynamically modulate these singularities. For instance, applying an external magnetic field to magneto-optical (MO) PhC slabs can lift the degeneracy of BIC modes and induces spin-orbit coupling, thereby resulting in intrinsically chiral BIC [22]. Additionally, the combined use of geometric perturbations and external fields allows precise tuning of polarization and topological charges, enabling arbitrary polarization states and intrinsic chirality [23]. Specifically, an external magnetic field breaks time-reversal symmetry (*TRS*) analogously to spatial symmetry breaking, shifting topological charges and altering polarization characteristics. Such dynamic control mechanisms have demonstrated considerable potential in condensed matter physics and photonics applications, including magnetically controlled vortex light encoders [24] and topological edge-state laser arrays [24-25].

BIC enable ultrahigh-Q optical resonances that support strong light–matter interactions, enhanced nonlinear responses, and extreme field localization, making them attractive for next-generation photonic applications such as ultrasensitive sensing, integrated lasing, and on-chip spectral manipulation. However, as application demands increase, intrinsic constraints of conventional BIC platforms become evident: ultrahigh-Q states typically occur only at high-symmetry points in the Brillouin zone, and the Q factor drops sharply once the system deviates from these symmetry-protected conditions. This reliance on momentum-space symmetry, together with limited tunability, constrains the implementation of robust, broadband, actively controlled, and reconfigurable photonic functionalities. To address these challenges, Moiré BIC has emerged as a promising alternative [26]. Formed by stacking two photonic crystal layers with slight misalignment, Moiré superlattices support relatively flat bands and sustain high-Q resonances across a much broader momentum space, substantially reducing sensitivity to fabrication imperfections and offering improved angular response and enhanced radiation suppression [26–28]. Geometric perturbations further enable precise control over topological singularities; for instance, laterally shifting one grating layer breaks $C_{2V}$ symmetry, leading to redistribution and splitting of topological charges and the emergence of new polarization states [32]. When combined with external stimuli such as magnetic fields, these structural modifications provide dynamic and versatile control over topological properties in Moiré photonic systems [33,34], enabling highly reconfigurable device functionalities for applications in sensing [35], light manipulation [36], and quantum optics [37].

In this work, we propose a reconfigurable Moiré BIC platform based on a two-dimensional MO PhC slab. As shown in Fig. 1, by applying external magnetic fields in various axis, the dynamic controllable splitting and shift of topological charges can be achieved. Specifically, we show that the lateral displacement of one grating layer breaks the $C_{2V}$ symmetry of the Moiré lattice, resulting in the generation of circularly polarized states with half-integer topological charges, denoted as $C$ points. Interestingly, the introduction of a magnetic field along the $z$ axis further allows the tuning of one of splitting $C$ points towards to the Γ point, leading to the formation of intrinsic chirality of Moiré BIC with distinct polarization properties. Importantly, our approach enables flexible, reversible control over the topology of Moiré BIC, offering a highly tunable system without the requirement to vary the geometric parameter of Moiré PhC structure. Our work opens exciting possibilities for designing reconfigurable Moiré chiral photonic devices, which leads to more applications in areas such as chiral sensing, chiral light manipulation, and quantum optics.

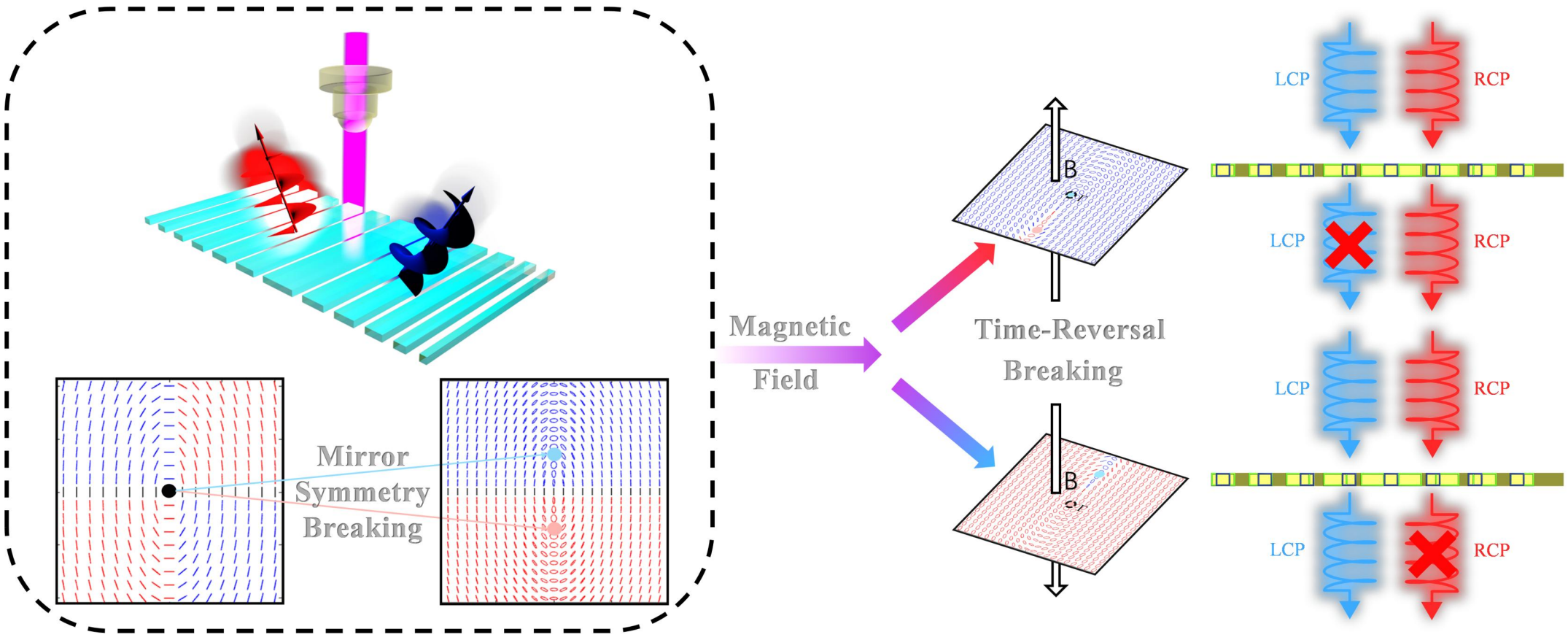

FIG.1 Schematic diagram of magnetic-field-controlled optical properties of Moiré gratings, illustrating how applying magnetic fields in different directions modulates the optical properties of the Moiré PhC, such as intrinsic chirality.

# II. RESULTS AND DISCUSSION

## A. Geometric Structure and Numerical Calculation

We construct a one-dimensional Moiré PhC slab composed of two parallel layers of Bismuth Iron Garnet (BIG) grating as illustrated in Fig. 2(a). The BIG is selected due to its strong optical transparency in the 400 nm to 2.5 μm wavelength range and pronounced MO response in the near-infrared region [38,39]. Its refractive index can vary between 3 and 3.5 by adjusting the doping concentration, and we fix its value at 3.4 in this work. Each layer of the Moiré PhC slab shares same thickness $d$ but have different periods and filling factors. Specifically, the upper grating has period $Λ_1$, and filling factor $F_1$, while the lower grating has period $Λ_2$ and filling factor $F_2$. To achieve a periodic Moiré superlattice, the grating periods should satisfy the commensurability condition: $Λ_2/Λ_1=N/(N+1)$, where $N$ is a positive integer defining the Moiré lattice unit size. Thus, one Moiré period comprises $N$ periods from the upper layer and $N$+1 periods from the lower layer, resulting in an overall lattice constant $Λ=Λ_1N=Λ_2(N+1)$. The geometric parameters of the photonic crystal are $d$=120nm, $Λ$=2990nm, $F_1$=0.36, $F_2$=0.66, and $N$=8, and the background medium is silica with a refractive index of 1.45. This geometry inherently preserves mirror symmetry along the $z$ axis ($σ_z$) and $C_{2V}$ rotational symmetry. Without external perturbations, these symmetries suppress the coupling between the bound states and radiative continuum, thereby enabling Moiré BIC. We investigate the behavior of half integer topological charges by systematically tuning geometric parameters and applying external magnetic fields, aiming to induce intrinsic and tunable chirality.

Our numerical analyses utilize the wave optics module in COMSOL Multiphysics, employing eigenfrequency solver as described previously [41]. We calculate eigenmodes and corresponding Stokes parameters to study the splitting and displacement of topological charges ($q$):

$$q=\frac{1}{2\pi}\int_{L} dk_{\parallel}\cdot\nabla_{k_{\parallel}}\psi\left(k_{\parallel}\right) \tag{1}$$

where $\psi$ ($k_{||}$) is the azimuthal angle of the polarization states' major axis, and $L$ is a closed path around the singularity.

Fig. 2(b) shows the calculated band structure, highlighting Moiré BIC band in red curve. Introducing a lateral displacement perturbation breaks $C_{2V}$ symmetry and shifts the bands near Γ point downward. Electric field distributions, shown in Fig. 2(c), confirm that fields remain localized and non-radiative at the Γ point (i.e., formation of Moiré BIC). In Fig. 2(c), the upper panel corresponds to the case without any lateral displacement, where the Moiré BIC remains strictly non-radiative at the Γ point, and the electric field is strongly localized inside the Moiré grating without free-space leakage. The lower panel of Fig. 2(c) shows the result after applying a 2 nm shift, where the broken symmetry allows radiation to occur in the off-Γ regions, indicating the emergence of quasi-BIC behavior due to the coupling of bound states to free space modes. Fig. 2(d) shows the numerical result of the Q factor with the variation of shifts. The existence of perturbation reduces the Q factor at the Γ point due to the breaking of $C_{2V}$ symmetry. Nevertheless, the off-Γ Q factors remain exceedingly high (above $10^5$), surpassing conventional BIC created by the standard gratings with a period of $A_2$ and a filling factor of $F_2$.

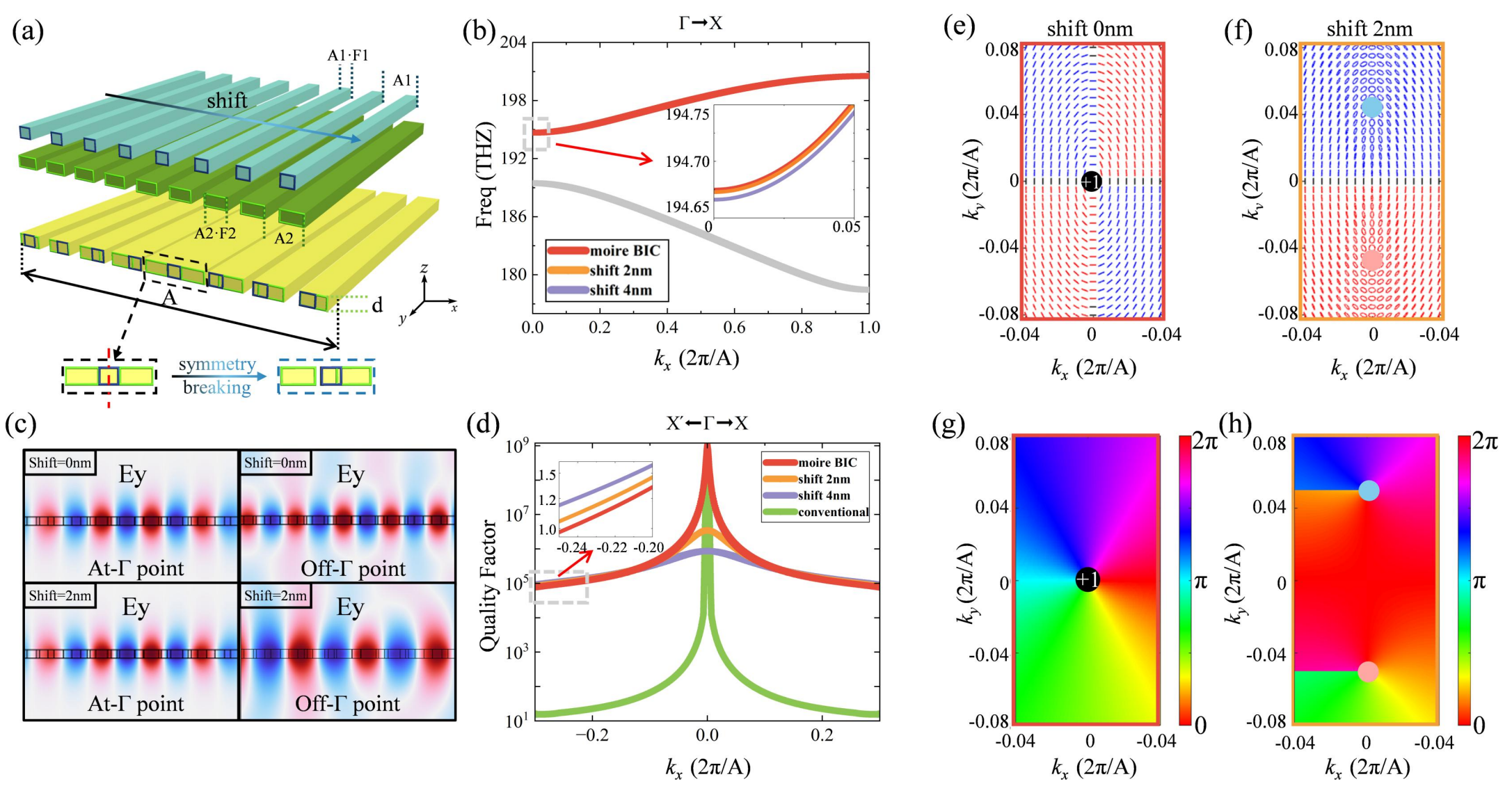

FIG. 2(a) Schematic diagram of the MO PhC slab structure, composed of two layers of gratings with different periods and duty cycles, which are merged into a single layer. Note that the gap between the single grating in the upper layer and the two gratings in the lower layer should be centrally aligned. The shift corresponds to a positive $x$ axis displacement of the upper blue grating. (b) Band structure calculation results for the MO PhC and the band structure calculation results after applying a shift to the upper layer of the grating. Due to the slight change in the geometric parameters, the band structures after the shift almost overlap. (c) $y$ direction electric field modes correspond to different characteristic frequency. The top image corresponds to the case without the shift displacement, and the bottom image corresponds to the electric field distribution after a 2 nm shift. (d) Q factor calculation results for the MO Moiré PhC and the MO conventional PhC at the same wavevector scale. (e)-(f) Far-field polarization diagrams correspond to different shift values. The +1 topological charge is marked with a small black sphere, and the blue and red spheres correspond to half topological charges with Stokes parameter S3<0 and S3>0, respectively. The red and blue ellipses represent right-handed and left-handed eigenpolarization states, respectively, and the black lines indicate the intrinsic linear polarization state. (g)-(h) Phase topological vortex diagrams in *k* space.

Unlike single layer gratings, which typically allow only one topological charge to block the zeroth-order diffraction channel, the Moiré grating folds the Brillouin zone into a smaller Moiré Brillouin zone, simultaneously supporting multiple topological charges. This significantly reduces scattering and in-plane propagation, enhancing radiative properties across the entire momentum space. To validate the existence and nature of symmetry protected Moiré BIC, we analyze the far-field polarization through the azimuthal angle and ellipticity angle from the Stokes parameters, as shown in Fig. 2(e). At the Γ point (marked by a black dot), the BIC is surrounded by linearly polarized radiation fields, with red indicating a positive S3 Stokes parameter, and blue indicating a negative S3. The topological charge of symmetry protected Moiré BIC at Γ is identified as +1. Introducing geometric perturbations splits this integer charge into two half integer charges ($q = \pm 1/2$), each exhibiting opposite circular polarization (C point) handedness, as shown in Figs. 2(e) and (f). This splitting occurs because symmetry reduction allows the originally degenerate state to evolve into two distinct states, represented in momentum space by pairs of *C* points marked by red and blue discs, indicating right-handed and left-handed circular polarizations, respectively. The decoupling of the original +1 topological charge into two half integer charges results in a redistribution of polarization and phase structure within momentum space. This effect can be clearly observed in the phase diagrams in Figs. 2(g) and (h), where topological vortex phase shifts indicate polarization property changes due to $C_{2V}$ symmetry breaking. These observations highlight how Moiré geometry enables control over topological charges and polarization characteristics via tuning geometric perturbations in the Moiré PhC system.

## B. Magnetically Control of Topological Singularities

In this section, we investigate how topological singularities evolve under external magnetic fields applied in different directions. In Moiré PhC systems, the distribution of topological charges strongly depends on the symmetry of the system. When an external magnetic field is introduced, it breaks the *TRS*, significantly altering both the topological charge and optical properties. Under these conditions, the optical response of MO PhC can be described by the permittivity tensor [44,45]:

$$\vec{\varepsilon} = \begin{pmatrix} \varepsilon & -iH_z & -iH_y \\ iH_z & \varepsilon & -iH_x \\ iH_y & iH_x & \varepsilon \end{pmatrix} \tag{2}$$

where the permittivity $\vec{\varepsilon}$ characterizes the material's response to the electric field, which is approximately 11.56. Here, $H_x$, $H_y$, and $H_z$ denote phenomenological coefficients describing the magnetization-induced gyration, which scale with the strength of the applied magnetic field rather than representing the physical magnetic field. The governing equation for the MO PhC slab can be expressed as follows [42]:

$$\nabla \times \mu_r^{-1}\left(\nabla \times \mathbf{E}_{xy}\right) - k_0^2\left(\vec{\varepsilon} - \frac{i\vec{\sigma}}{\omega\vec{\varepsilon}}\right)\mathbf{E}_{xy} = 0 \tag{3}$$

where $\sigma$ is the electrical conductivity of the material (set to 0), $\mu_r$ is the relative magnetic permeability (set to 1), and $i$ is the imaginary unit. These parameters critically influence the electromagnetic behavior of the Moiré PhC system. The MO effect has a significant impact on the behavior of the Moiré system. Applying an external magnetic field to a periodically modulated MO material induces nonreciprocity in its permittivity tensor, breaking the system's *TRS*, which induces spin orbit coupling effects in the photonic Bloch waves within momentum space. As the magnetic field strength further increases, it continuously modulates the Berry curvature distribution in momentum space [46], thereby enabling the controlled evolution of topological singularities associated with the photonic bands. We performed numerical simulation to quantify these effects. Figs. 3(a) and (b) show the calculated Q factor under different magnetic field orientations. Introducing a magnetic field reduces the Q factor due to symmetry breaking along specific axis. A magnetic field applied along the $x$ axis breaks the in-plane mirror symmetry, inducing TE-TM mode hybridization and non-reciprocal radiative losses. Conversely, a magnetic field applied along the $z$ axis breaks the out of plane mirror symmetry, perturbs the Berry curvature distribution, and amplifies group velocity dispersion and diffraction effects, Which further opens radiative channels, transforming the Moiré BIC into a quasi-BIC coupled to free space radiation. Increasing the magnetic field intensity further enhance these leakage modes, progressively reducing the Q factor.

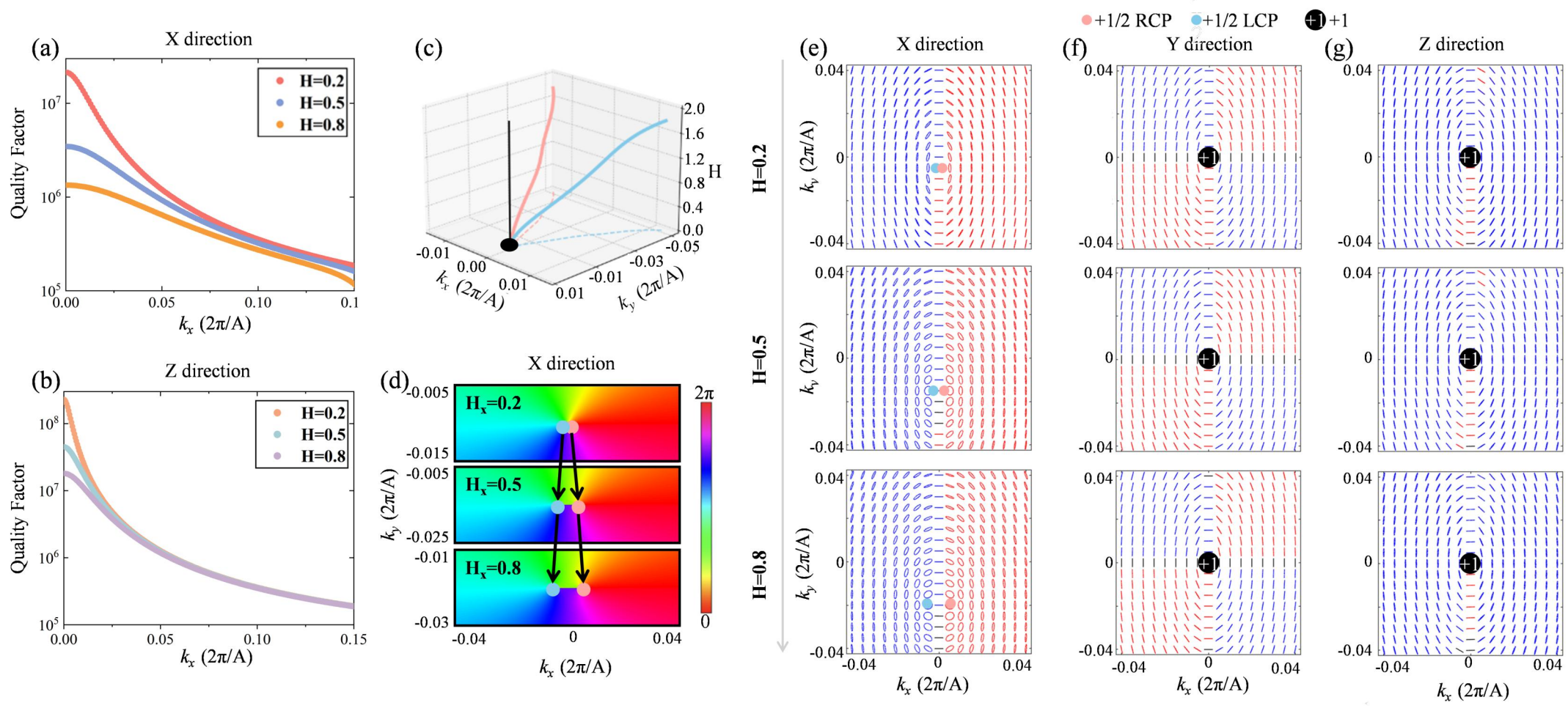

FIG. 3(a)-(b) Q value calculation results under the influence of magnetic fields in different directions. (c) Schematic diagram of the trajectory of the topological charge in momentum space as the magnetic field strength increases. The blue and red lines correspond to the movement trajectories of half topological charges with different S3 parameters under the $x$ axis magnetic field, while the black line corresponds to the movement trajectory of the +1 integer topological charge under the influence of $y$ axis and $z$ axis magnetic fields. (d) Phase diagram of the topological vortex in $k$ space under different $x$ axis magnetic field strengths. The black arrows indicate the motion trajectory. (e)-(g) Far-field polarization diagrams under the influence of magnetic fields in different directions.

As shown in Fig. 3(c), applying a magnetic field along the $x$ axis partially breaks the original $C_{2V}$ symmetry of the system. Specifically, the mirror symmetry along the $y$ axis ($\sigma_y$) is preserved, while symmetry along the $x$ axis ($\sigma_x$) is disrupted. This selective symmetry breaking causes the initially bound integer topological charge ($q$=1) at the Γ point to split into two half-integer topological charges ($q$=1/2).

Additionally, the magnetic field component $H_x$ introduces a non-reciprocal gradient of the Berry curvature gradient in $k$ space, acting as an effective $k$ space force toward the negative $k_y$ direction. This force uniformly couples to both half integer vortices, collectively translating the pair in the negative $k_y$ direction and simultaneously increasing their separation along both $k_x$ and $k_y$ directions as the magnetic field strength $H_x$ grows. The trajectory of these topological charges, shown in Fig. 3(c), reveals a nearly linear $C$ point's displacement in $k$ space with increasing magnetic field strength $H_x$. The corresponding far field polarization phase diagram presented in Fig. 3(d) clearly demonstrates this movement of the topological charges, which further confirms the motion of topological charges across $k$ space. In contrast, as shown in Fig. 3(f), applying a magnetic field along the $y$ axis maintains the system's $\sigma_x$ symmetry, thereby restricting coupling between TE and TM modes. The induced vector potential from the $y$ axis magnetic field lies parallel to the plane of PhC, introducing only longitudinal phase modulations without altering the relative phase between the transverse polarization components. Consequently, the distribution of the Stokes parameter $S_3$, both locally and globally remains unchanged. Finally, as shown in Fig. 3(g), when a magnetic field is applied in the $z$ axis, it breaks the *TRS* but preserves the $C_{2V}$ symmetry. Therefore, the topological charge remains fixed at the Γ point with no displacement. At moderate field strengths, $k$ space polarization states do not significantly exhibit circular polarization. However, increasing the $z$ axis magnetic field $H_z$ gradually biases the polarization states uniformly toward a single circular polarization component, causing the Stokes parameter $S_3$ to become uniformly negative throughout momentum space converges uniformly to negative values.

To thoroughly investigate the phenomena induced by the MO effect on the Moiré BIC, we introduced a lateral grating displacement of 0.5 nm, breaking the $C_{2v}$ symmetry of the Moiré PhC. This placement splits the integer topological charge originally located at the center of $k$ space into two chiral half integer topological charges. We then simulated the system's response to magnetic fields of varying strengths in different directions. Fig. 4(c) shows how the two-half integer topological charges evolve in $k$ space under an increasing magnetic field oriented along the $x$ axis. As the filed strength rises, these half integer topological charges gradually separate and shift collectively along the $k_y$ direction. Fig. 4(a), depicted by the red curve, shows the polarization state evolution at the Γ point under the $x$ axis magnetic field. Here, the polarization transition smoothly from vertically linear polarization to near-circular polarization, ultimately approaching at the Γ point gradually transitions from a and eventually shifts toward a horizontally linear polarization. Fig. 4(f) displays the trajectory of topological singularities in $k$ space, clearly demonstrating an approximately linear shift with increasing field strength. The observed linear separation and the synchronized drift of these half-integer changes are predominantly drive by symmetrical constraints and perturbation in weak magnetic fields.

Fig. 4(d) shows the far-field polarization state after applying a magnetic field along the $y$ axis. In this configuration, the initial lateral displacement breaks the $C_{2V}$ symmetry, and subsequent magnetic field application along the $y$ axis further disrupts both *TRS* and mirror symmetries. This additional symmetry breaking introduces nonreciprocal effects, stabilizing the half integer topological charges in $k$ space without noticeable positional shifts, even as the magnetic field strength increases. Significantly, the lateral displacement of the grating introduces symmetry breaking that creates additional degrees of freedom, enabling effective magnetic field control of circular polarization $C$ points.

When applying a magnetic field along the *z* axis, both the *TRS* and out of plane mirror symmetry ($\sigma_z$) are broken, which generates a nontrivial Berry curvature distribution aligned with the external magnetic field, promoting the migration of topological charges. As shown in Figs. 4(e) and (g), the pair of *C* points move closer with increasing magnetic field strength, ultimately forming intrinsic chiral Moiré BIC states. The blue curve in Fig. 4(a) highlights how polarization at the Γ point evolves under a *x* axis magnetic field, transitioning from vertical linear polarization to LCP with stronger fields.

It is noteworthy that, compared to previous studies, shifting topological charges within a Moiré lattice generally demands stronger magnetic fields. This increased requirement arises because Moiré interleaving folds the Brillouin zone, reducing the curvature radius of Moiré bands and thereby enhancing the effective robustness of topological charges. Consequently, larger magnetic fields are necessary to move the *C* points toward the center of Brillouin-zone. Therefore, employing a smaller displacement, for example 0.5 nm, will effectively constrain the separation of the *C* points, enabling intrinsic chirality under moderate magnetic field strengths. In the subsequent analysis, we maintain this 0.5 nm shift to systematically investigate these effects in detail.

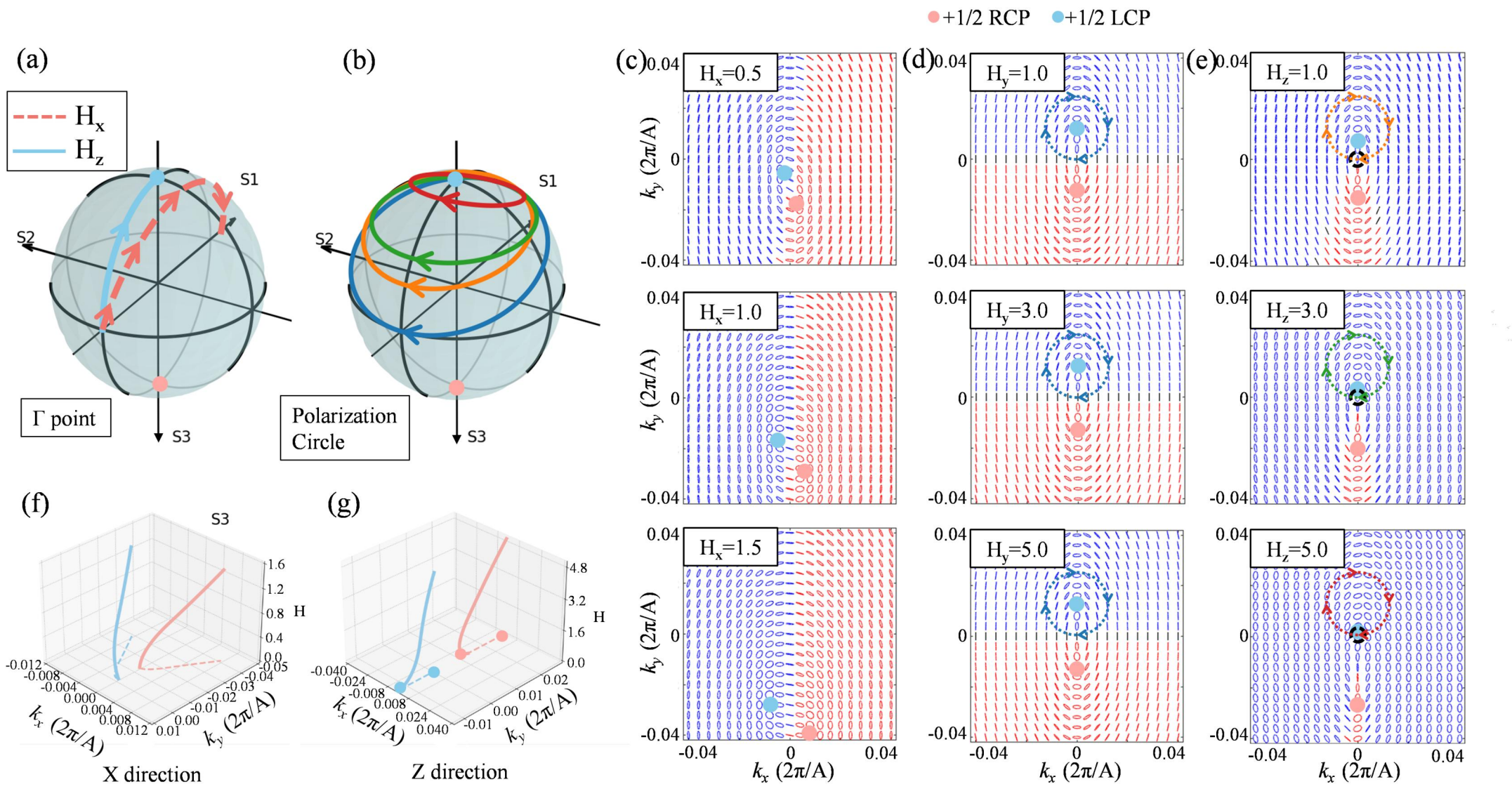

FIG. 4(a) The polarization state evolution trajectory at the Γ point on the Poincaré sphere, where the red dashed line corresponds to the trajectory under the application of a magnetic field along the *x* axis, and the blue solid line represents the trajectory under the magnetic field along the *z* axis. (b)The closed loop in *k* space is represented on the Poincaré sphere as the corresponding polarization evolution, where the different colors indicate the polarization states under different magnitudes of the magnetic field applied along the *z* axis. (c)-(e) *k* space far-field polarization under the influence of magnetic fields along different directions when the shift is 0.5 nm. In the *x* axis, as the magnetic field strength increases, the splitting in the $k_x$ direction becomes larger, and both topological charges move together in the negative $k_y$ direction. In the *z* axis, as the magnetic field strength increases, the two half topological charges move together in the negative $k_y$ direction. (f)-(g) Movement trajectories of the topological charge as the magnetic field strength changes under different magnetic field strengths.

By adjusting the external magnetic field along the *z* axis, we achieve an intrinsically chiral Moiré BIC. To further investigate how topological charges evolve within the band structure, we calculated the 3D band diagrams of the MO Moiré PhC. As shown in Figs. 5(a), 5(e), and 5(h), the topological charges on the band structure are marked by blue and red points, which move across the band as the magnetic field changes. Specifically, as shown in Figs. 5(b) and 5(i), at magnetic field strengths of $H_z$= $\pm$5.3, one of the *C* points migrates to the Γ point, indicating the formation of intrinsic chiral Moiré

BIC. It is important to highlight those previous approaches to generating chiral BIC typically required simultaneous breaking of $C_{2v}$ and $\sigma_z$ symmetries. Such approaches restricted chiral emission to only one side of the photonic nanostructure and limited the quality factor of intrinsic chiral BIC [47]. In contrast, our approach utilizes an external magnetic field to break *TRS* dynamically, enabling intrinsic chiral BIC that exhibits chirality on both sides of the Moiré PhC slab. This enhances light-matter interaction, opening new opportunities for the development of high-performance chiral lasers [48-50].

Figures. 5(c) and 5(d) show simulated transmission spectra under LCP and RCP incidences, respectively. From the spectra, it can be observed that for RCP incidence, there is a positive *C* point charge marked by red dot in the transmission spectrum, while LCP incident waves can excite the intrinsic state at the Γ point, indicating the presence of a *C* point with Left-handed chirality, corresponding to Fig. 5(b). Fig. 5(g) compares transmission spectra under RCP (red curve) and LCP (blue curve) incidences at different magnetic field strengths. We calculated the corresponding circular dichroism (CD) at different wavelengths using the established formula [51,52]:

$$\mathrm{CD} = \frac{T_{\mathrm{LCP}} - T_{\mathrm{RCP}}}{T_{\mathrm{LCP}} + T_{\mathrm{RCP}}} \tag{4}$$

$T_{\mathrm{LCP}}$ and $T_{\mathrm{RCP}}$ denote the transmittance under LCP and RCP incidence, respectively. The CD value ranges from -1 to +1, where a positive value indicates stronger transmission for LCP, while a negative value indicates stronger transmission for RCP. At the wavelength corresponding to the Γ point, the CD exceeds 0.99, confirming the strong intrinsic chirality of the Moiré BIC.

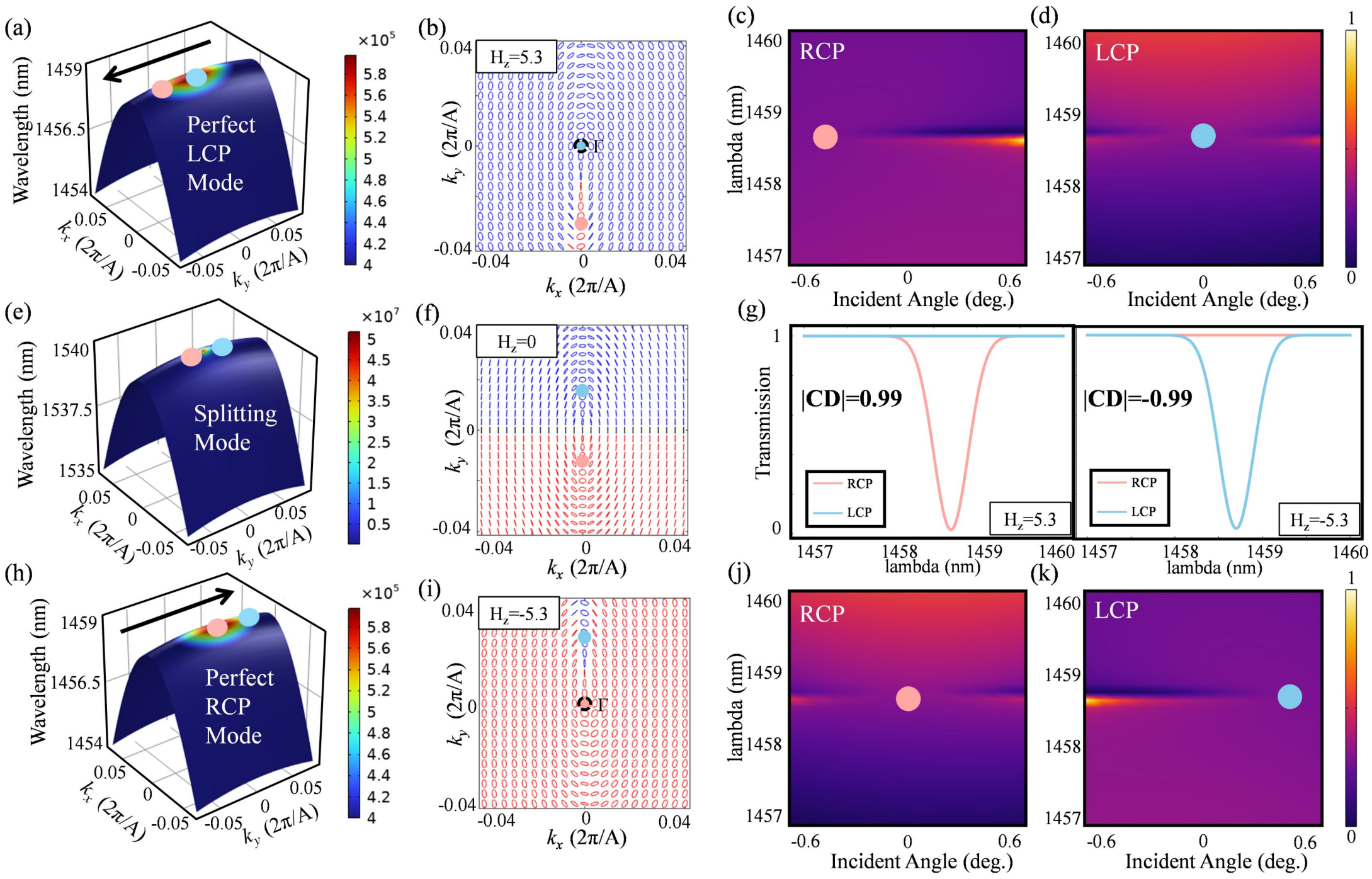

FIG. 5(a) Band structure and Q factor calculation results when $H_z = 5.3$. The red and blue dots represent the position of the *C* point on the band, with the blue corresponding to the LCP intrinsic mode.(b) Far-field polarization at $H_z = 5.3$, with the $\Gamma$ point marked by a black dashed circle.(c)-(d) Corresponding transmission spectra, with the *C* point positions marked by red and blue dots.(e) Topological charge splitting mode due to the breaking of $C_{2v}$ symmetry when no magnetic field is applied.(f) Far-field

polarization diagram when no magnetic field is applied.(g) Transmission spectra for vertical incidence under two different magnetic field strengths. The left side corresponds to the LCP intrinsic mode, and the right side corresponds to the RCP intrinsic mode. (h) Band structure and Q factor calculation results when $H_z =- 5.3$. The red and blue dots represent the position of the *C* point on the band, with the red corresponding to the RCP intrinsic mode. (i) Far-field polarization at $H_z =- 5.3$, with the Γ point marked by a black dashed circle. (j)-(k) Corresponding transmission spectra, with the *C* point positions marked by red and blue dots.

Finally, we investigate the tunability of intrinsic chiral BIC under external magnetic field. As illustrated in Fig. 5(h), applying a downward-directed magnetic field ($H_z$=-5.3) drives the RCP associated *C* point (marked by a red dot) toward the Γ point, again producing of an intrinsic chiral Moiré BIC. The associated the transmission spectra under LCP and RCP incidences at $H_z$=-5.3 are shown in Figs. 5(j) and (k). The disappearance points corresponding to the pair of *C* points in these spectra (indicated by blue and red dots) are in excellent agreement with the distribution of topological singularities shown in Fig. 5(i). Therefore, reversing the magnetic field axis allows the intrinsic chiral BIC to be dynamically reconfigured, demonstrating effective control over optical chirality.

# III. CONCLUSIONS

In this work, we propose and numerically demonstrate a two-dimensional MO Moiré PhC slab that supports magnetically tunable BIC and their associated topological singularities. The structure is formed by stacking two diffraction gratings with mismatched periods into a Moiré grating exhibited $C_{2V}$ symmetry. In the absence of perturbations, an integer vortex carrying a topological charge of +1 is stabilized at the Γ point. Upon applying an external magnetic field, we systematically analyze the directional dependence of the external magnetic field resulting phase singularities in *k* space. Breaking the $C_{2V}$ symmetry by introducing a horizontal displacement between the layers gives rise to a pair of *C* points with opposite chirality. Subsequent application of a magnetic field along the *z* axis breaks *TRS*, resulting in an intrinsically chiral Moiré BIC with a CD value from -0.99 to 0.99. Our model integrates Moiré geometric modulation with tunable MO coupling while preserving the underlying PhC geometry. By jointly adjusting the lateral and the direction and magnitude of the external magnetic field, we achieve controlled the splitting of the topological charge from integer to half integer values, directional drift of singularities in *k* space, and reversible tuning of chiral BIC. These magnetically controlled Moiré BIC exhibit ultrahigh Q value and robust chirality over a broad range of incidence angles, providing a promising platform for topological lasers, nonreciprocal optical isolators, and chiral photonic sensors.

# ACKNOWLEGEMENTS

This work is supported by the National Natural Science Foundation of China (Grants No. 62104174, No. 62205253 and No. 92373102), and Hubei Provincial Natural Science Foundation of China (Grants No. 2021CFB054). The authors would like to acknowledge Prof. Xiaoze Liu, Dr. Tianle Ke and Dr. Junyu Li for their valuable discussion.

## APPENDIX A: MAGNETO OPTICAL EFFECT IN CONVENTIONAL GRATINGS

To further illustrate the unique topological behavior of the Moiré PhC, we investigate the MO response of a conventional single-layer grating composed of the same BIG material. Unlike the Moiré structure, the conventional grating exhibits uniform periodicity and higher spatial symmetry. When an external magnetic field is applied along the *x* axis, *TRS* is broken, enabling the formation and manipulation of topological charges in momentum space. The band structure of the single-layer grating, shown in Fig. 6(a), reveals an upward shift in resonance frequencies with increasing magnetic field strength. Corresponding Q factor results in Fig. 6(b) indicate a sharp decrease at the Γ point as $C_2$ symmetry is progressively broken. Furthermore, the Q factor decays more rapidly across the momentum space compared to the Moiré grating, indicating reduced spatial robustness. While both the Moiré grating and the conventional grating exhibit topological charge splitting along the $k_x$ axis under a *x*-directed magnetic field, their evolution trajectories in momentum space differ markedly. In the Moiré superlattice structure, as shown in Fig. 3(e), the pair of split half-integer charges collectively drift toward the negative $k_y$ axis, forming a downward trajectory. In contrast, in the conventional grating, as shown in Figs. 6(c) and 6(f), the split charges migrate toward the positive $k_y$ axis under the same magnetic conditions. This contrasting behavior highlights the critical role of Moiré-induced spatial symmetry modulation and interlayer interference in shaping the dynamics of topological singularities. The Moiré superlattice not only influences the axis of charge splitting but also governs the global trajectory of topological features in momentum space. These findings demonstrate that, while MO effects are present in both structures, the engineered geometric phase and broken symmetry in Moiré gratings enable a level of topological control that is unattainable in conventional single-layer designs.

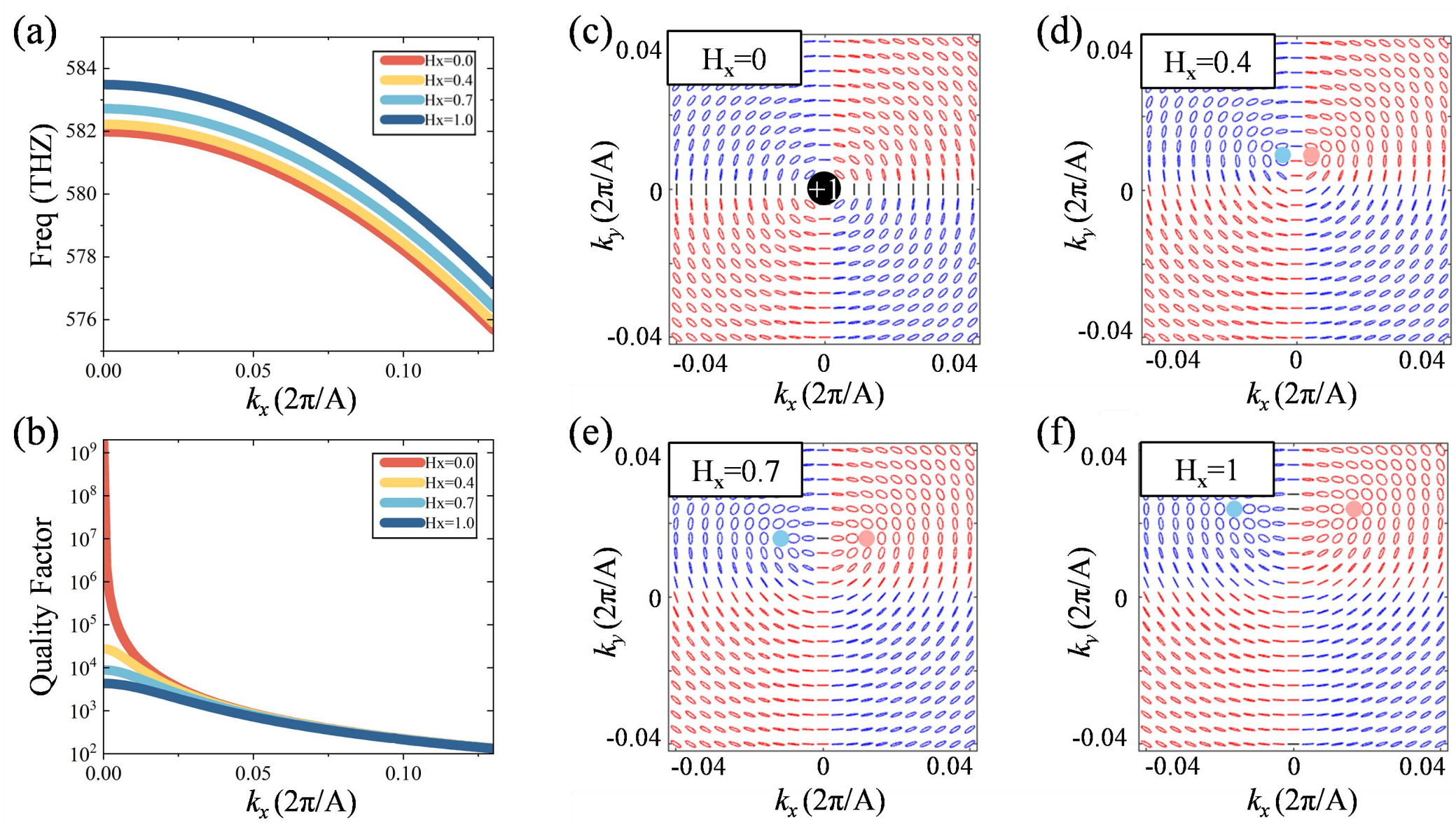

FIG. 6(a) Band structure calculation results for the single-layer grating under varying strengths of the magnetic field applied along the *x* axis. (b) Q factor calculation results for the single-layer grating under varying strengths of the magnetic field applied along the *x* axis. (c) - (f) Far-field polarization maps under varying magnetic field strengths along the *x* axis, the topological charge is marked with black dots, *C* points for LCP with blue dots, and *C* points for RCP with red dots.

## APPENDIX B: IMPACT OF GEOMETRIC PARAMETERS ON MOIRÉ BOUND STATES IN THE CONTINUUM

To investigate the structural tunability of Moiré BIC, we systematically studied the influence of key geometric parameters, particularly the Moiré superlattice commensurability ratio $N$ and the grating thickness $d$, on the photonic band structure and associated Q factors.

Fig. 7(a) and 7(b) present the calculated band structures for $N = 4$ and $N = 6$, respectively. As $N$ decreases, the Moiré supercell becomes smaller, resulting in a weaker Brillouin zone folding effect. This leads to increased radiative coupling and reduced mode confinement, causing a downward shift in resonance frequencies and a significant degradation in Q factors. Figure 7(c) shows the corresponding Q factor distributions, where the blue curve represents $N = 6$ and the green curve represents $N = 4$. The results indicate that selecting an optimal $N$ enables the formation of flatter bands and stronger confinement, thereby supporting higher-Q resonant states.

We also examined the influence of grating thickness on BIC performance. Figs. 7(d) and (e) depict the band structures for structures with thicknesses $d$=126 nm and $d$=136 nm, respectively. Increasing the thickness extends the optical cavity in the out-of-plane axis, which enhances radiative leakage. As a result, resonance frequencies exhibit a slight redshift, and the corresponding Q factor experience a moderate reduction. Fig. 7(f) compares the Q factor distributions for the two thicknesses, with the orange curve representing $d$=126 nm and the purple curve representing $d$=136 nm. The data confirm that greater thickness leads to increased coupling with the radiation continuum, thus suppressing Q factors.

These results underscore the critical importance of geometric design in tuning the formation, spectral positioning, and radiative robustness of Moiré BIC. By carefully optimizing parameters such as the commensurability ratio $N$ and grating thickness $d$, the optical properties of the system can be flexibly engineered. This tunability provides a structural foundation for realizing high-Q topological modes and stable polarization singularities in practical photonic devices.

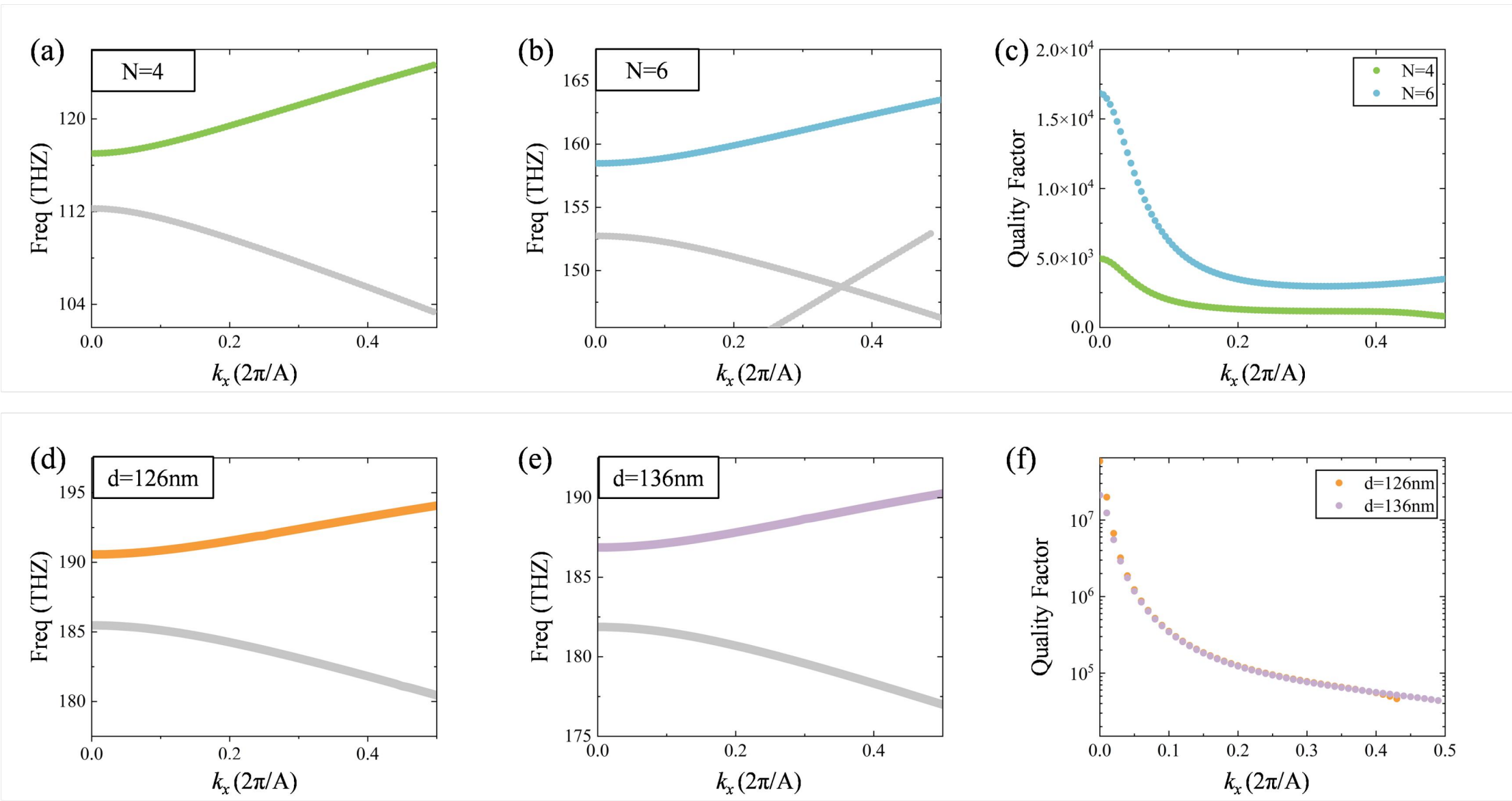

FIG. 7(a) Band structure calculation for the Moiré photonic crystal with commensurability ratio N = 4. The reduced supercell size leads to weaker Brillouin zone folding and lower mode confinement. (b) Band structure for N = 6, showing enhanced band folding and improved confinement, supporting higher-Q BIC modes. (c) Q factor distributions corresponding to (a) and (b). The blue curve represents N = 6, while the green curve corresponds to N = 4; higher Q values are observed for larger N.(d) Band structure of the Moiré photonic crystal with grating thickness d=126 nm.(e) Band structure for increased thickness d=136 nm,

showing a slight redshift in frequency and changes in mode confinement.(f) Q factor distributions corresponding to (d) and (e). The orange curve corresponds to d=126 nm, and the purple curve corresponds to d=136 nm, increased thickness results in moderate Q suppression due to enhanced radiative leakage.